\documentclass[letterpaper]{article} 
\usepackage{aaai24}  
\usepackage{times}  
\usepackage{helvet}  
\usepackage{courier}  
\usepackage[hyphens]{url}  
\usepackage{graphicx} 
\usepackage{natbib}  
\usepackage{caption} 
\usepackage{algorithm}
\usepackage{algorithmic}

\usepackage{newfloat}
\usepackage{listings}
\DeclareCaptionStyle{ruled}{labelfont=normalfont,labelsep=colon,strut=off} 
\floatstyle{ruled}
\newfloat{listing}{tb}{lst}{}
\floatname{listing}{Listing}
\usepackage{microtype}
\usepackage{amsmath}
\usepackage{graphicx}
\usepackage{subfigure}
\usepackage{booktabs}
\usepackage{multirow}
\usepackage{amsmath}
\usepackage{dsfont}
\usepackage{bbding}
\usepackage{pifont}
\usepackage{amsfonts,amssymb, utfsym}

\title{Token-Level Contrastive Learning with Modality-Aware Prompting for Multimodal Intent Recognition}
\author{
    Qianrui Zhou\textsuperscript{\rm 1,\rm 2}, 
    Hua Xu\textsuperscript{\rm 1,\rm 2}\footnote{Hua Xu is the corresponding author.}, 
    Hao Li\textsuperscript{\rm 1,\rm 2}, 
    Hanlei Zhang\textsuperscript{\rm 1,\rm 2},
    Xiaohan Zhang\textsuperscript{\rm 1, \rm 3}, 
    Yifan Wang\textsuperscript{\rm 1, \rm 3}, 
    Kai Gao\textsuperscript{\rm 3}
}
\affiliations{
    \textsuperscript{\rm 1}Department of Computer Science and Technology, Tsinghua University, Beijing 100084, China\\
    \textsuperscript{\rm 2}Beijing National Research Center for Information Science and Technology (BNRist), Beijing 100084, China\\
    \textsuperscript{\rm 3}School of Information Science and Engineering, Hebei University of Science and Technology, Shijiazhuang 050018, China\\

    zgr22@mails.tsinghua.edu.cn, xuhua@tsinghua.edu.cn

}

\begin{document}

\maketitle

\begin{abstract}
Multimodal intent recognition aims to leverage diverse modalities such as expressions, body movements and tone of speech to comprehend user's intent, constituting a critical task for understanding human language and behavior in real-world multimodal scenarios. Nevertheless, the majority of existing methods ignore potential correlations among different modalities and own limitations in effectively learning semantic features from nonverbal modalities. In this paper, we introduce a token-level contrastive learning method with modality-aware prompting (TCL-MAP) to address the above challenges. To establish an optimal multimodal semantic environment for text modality, we develop a modality-aware prompting module (MAP), which effectively aligns and fuses features from text, video and audio modalities with similarity-based modality alignment and cross-modality attention mechanism. Based on the modality-aware prompt and ground truth labels, the proposed token-level contrastive learning framework (TCL) constructs augmented samples and employs NT-Xent loss on the label token. Specifically, TCL capitalizes on the optimal textual semantic insights derived from intent labels to guide the learning processes of other modalities in return. Extensive experiments show that our method achieves remarkable improvements compared to state-of-the-art methods. Additionally, ablation analyses demonstrate the superiority of the modality-aware prompt over the handcrafted prompt, which holds substantial significance for multimodal prompt learning. The codes are released at \url{https://github.com/thuiar/TCL-MAP}.

\end{abstract}

\section{Introduction}

Intent recognition is a pivotal component in natural language understanding(NLU), which is employed to classify intent categories in goal-oriented scenarios based on text information. Prior studies have extensively researched intent recognition and validated the significance of textual modality \cite{zhang2021textoir, zhang2021discovering}. Recently, multimodal intent recognition has been devised to analyze human intents by incorporating both natural language and other nonverbal information (e.g. video and audio). Contrary to depending on single modality, leveraging multiple modalities can provide a substantial amount of information, providing a distinct advantage in accurately identifying more intricate intent categories. Pioneering works in this field \cite{10.1145/3503161.3547906, saha-etal-2020-towards} gather multimodal data from the real world to create intent recognition datasets, which affirms the crucial role of multimodal information on intent recognition tasks.

To effectively leaverage the data from various modalities, numerous methods have been proposed for multimodal language understanding. As the state-of-the-art methods for multimodal intent recognition, \cite{tsai2019multimodal, 10.1145/3394171.3413678, rahman2020integrating} utilize Transformer-based techniques to integrate information from different modalities into a unified feature. Another series of works primarily focus on achieving significant representations by addressing the correlations among modalities. For instance, \cite{dong2022improving} introduce a framework for cross-modality contrastive learning aimed at narrowing the gap between modalities and enhancing multimodal representations. Moreover, the latest work \cite{yu-etal-2023-speech} aims to incorporate knowledge from large-scale pre-trained models and has demonstrated remarkable progress. 

However, existing approaches are adept at capturing basic semantics such as emotions and encounter limitations in capturing the deep semantic information inherent in intents using multimodal data, owing to deficiencies in managing potential correlations across distinct modalities and extracting high-quality semantic features from nonverbal modalities.To achieve a deeper understanding of human intents in the real world, there are two main challenges. Firstly, given that intent recognition is primarily a text-centric task, it is crucial to explore the association between nonverbal modalities and text modality. Secondly, mining semantic information from video and audio modality poses a significant difficulty due to the complex nature of the intent concept.

\begin{figure*}[h!]
	\centering  
	\includegraphics[scale=.21]{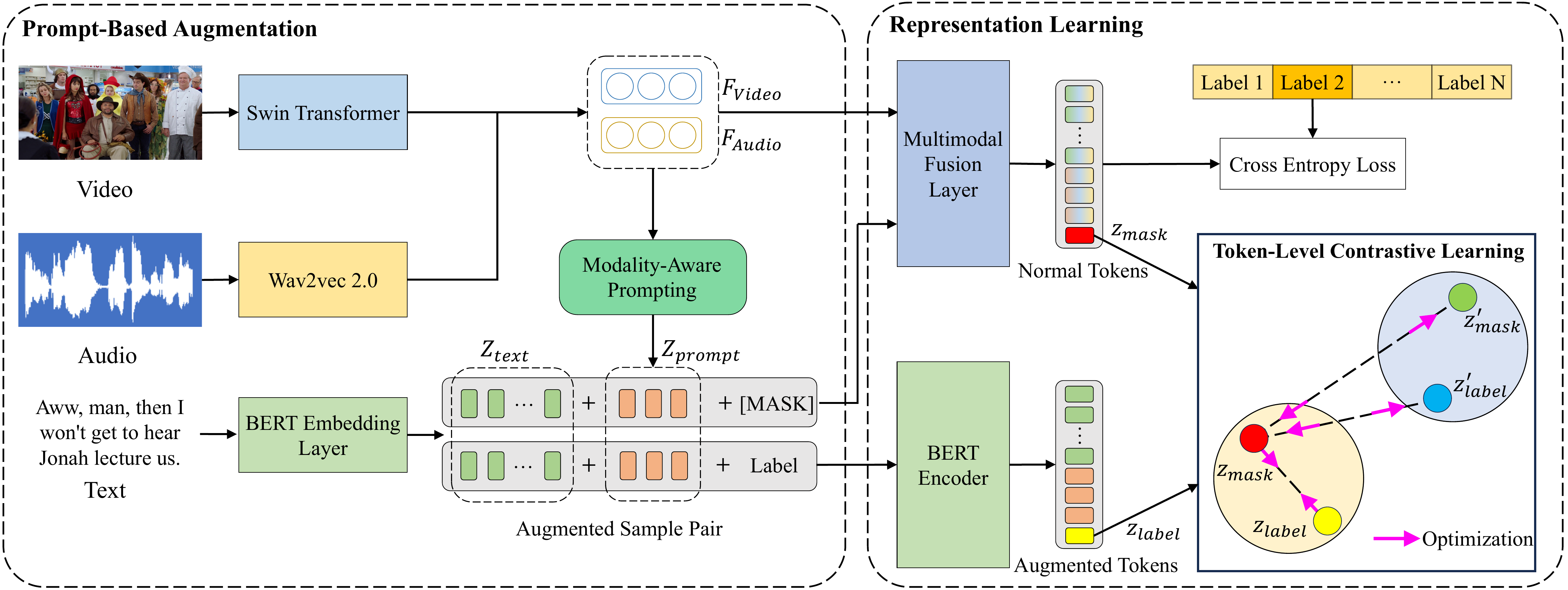}
	\caption{ \protect \label{TCL_MAP} The overview architecture of TCL-MAP. In the Prompt-Based Augmentation module, we first create the modality-aware prompt using multimodal features, and then concatenate text tokens, prompt tokens and [MASK]/Label token to construct augmented pair. In the Representation Learning module, we extract the refined tokens for classification and conduct  contrastive learning between the [MASK] token and the Label token.
        }
\end{figure*}

In this paper, we propose a Token-Level Contrastive Learning with Modality-Aware Prompting (TCL-MAP) approach, which generates prompts based on video and audio modalities to enhance the text representation and in return utilizes the high-quality textual feature to guide other modalities in learning semantic features. To tackle the first challenge, we design a modality-aware prompting (MAP) module to align different modalities based on their similarity and generate modality-aware prompt using a cross-modality attention mechanism. The generated prompt combines video and audio information with learnable parameters to establish a foundational semantic environment for text representation learning, mitigating the need for labor-intensive prompt engineering. For the second challenge, we utilize a token-level contrastive learning framework (TCL) which leverages premium supervised signals to facilitate the extraction of semantic information. Specifically, TCL initially constructs augmented samples by concatenating the modality-aware prompt and ground truth labels with the original text. Subsequently, TCL extracts the ground truth token from BERT \cite{devlin2018bert} to guide the process of semantic feature learning through the application of NT-Xent \cite{sohn2016improved} loss.

Extensive experiments are performed on MIntRec \cite{10.1145/3503161.3547906} and MELD-DA \cite{saha-etal-2020-towards} datasets, which demonstrates the significant improvements achieved by our methods over the state-of-the-art methods. Moreover, ablation analyses reveal that the modality-aware prompt generated by MAP outperforms the handcrafted prompt, thereby contributing to the advancement of multimodal prompt learning. Our contributions can be summarized as follows:

\begin{itemize}
    \item We design a modality-aware prompting module to generate modality-aware prompts using video and audio modalities, which establishes an optimal multimodal semantic context for text modality. 

    \item We propose a token-level contrastive learning framework for leveraging semantic information from the ground truth label to guide other modalities in learning semantic representations. To the best of our knowledge, this is the first attempt that utilizes prompt learning to create premium supervised signals for contrastive learning.

    \item Comprehensive experiments conducted on two challenging datasets show that the proposed method achieves state-of-the-art performance on the multimodal intent recognition task.
\end{itemize}

\section{Related Works}
\subsection{Multimodal Fusion Methods}
Multimodal fusion techniques strive to achieve high-quality multimodal representations through effective fusion processes. Traditional approaches \cite{zadeh2017tensor, liu2018efficient, hou2019deep} harnesses the representational capabilities of tensors for multimodal data representation, showcasing their representational capabilities. However, these methods face the challenge of striking a balance between computational complexity and the quality of tensor representation. To solve the problem, MFN \cite{zadeh2018memory} first learns view-specific interactions and leverages an attention mechanism with a multi-perspective gated memory.

Recent methods are based on Transformer for multimodal fusion. By incorporating attention mechanism, MulT \cite{tsai2019multimodal} addresses the challenge of non-aligned multi-modal sequences and MISA \cite{10.1145/3394171.3413678} aims to learn representations that exhibit both modality-invariant and modality-specific characteristics. Moreover, MAG-BERT \cite{rahman2020integrating} introduces an attachment mechanism to enhance the fine-tuning process of BERT \cite{devlin2018bert}. Subsequently, researchers start to prioritize the interactions within individual modalities. For example, MMIM \cite{han2021improving} focuses on the correlations between unimodal inputs and multimodal fusion features through maximizing the mutual information. To separately address the relationships between each pair of modalities, BBFN \cite{han2021bi} integrates two bi-modal fusion modules along with a gated control mechanism. Given that top-down interactions remain unaccounted in previous approaches, MMLatch \cite{paraskevopoulos2022mmlatch} addresses this limitation by incorporating a feedback mechanism in the forward pass.

\subsection{Prompt Learning}
Prompt learning originates from nature language processing (NLP) and is adopted to elicit useful information from pre-trained language models for downstream tasks. CoOp \cite{zhou2022learning} first employ prompt learning in adapting large vision-language models, garnering the interest of numerous researchers. Subsequently, CoCoOp \cite{zhou2022conditional} addresses the weak generalizability issue of CoOp by generating an input-conditional token for each sample. To further incorporate contextual information into the prompt, DenseCLIP \cite{rao2022denseclip} introduces a transformer encoder within the CoCoOp framework to enhance the correlation between image embeddings and prompt tokens. Recent works \cite{Wang_2022_CVPR, Li_2022_CVPR, gan2023decorate} have applied prompt learning to various domains in computer vision. Different from prior approaches, our method pioneers the application of prompt learning on multimodal tasks without reliance on extensive pre-trained models. Moreover, our research demonstrates the substantial impact of prompt learning in enhancing multimodal representations.

\subsection{Contrastive Learning}
Contrastive Learning emphasizes similarities and differences between data pairs, pulling similar pairs closer and pushing dissimilar pairs apart.
Early approaches  \cite{wu2018unsupervised, ye2019unsupervised, tian2020contrastive} lays the diverse foundation for contrastive learning's evolution, encompassing various models, loss functions, and pretext tasks.
Subsequently, MoCo \cite{he2020momentum} views contrastive learning as dictionary look-up and exploits a dynamic dictionary  with a queue and a momentum encoder. Notably, SimCLR \cite{chen2020simple} proposes a simple framework for contrastive learning using diverse data augmentation operations and a nonlinear projection head.

Recent methodologies aim to eliminate the necessity for negative examples and improve the efficacy of feature extraction. To enhance robustness in the presence of diverse augmentations, BYOL \cite{grill2020bootstrap} proposes a slow-moving average target network output using the online network's output, effectively avoiding the need for negative pairs. Taking a step further, SimSiam \cite{chen2021exploring} discards negative pairs and the momentum encoder, relying on identical encoders and stop-gradient mechanism to prevent output collapse. Besides, DINO \cite{caron2021emerging} employs vision transformers as the foundation for self-supervised learning, implementing self-distillation without requiring any labels.

\section{Method}
\subsection{Overview}
In this section, we describe the architecture of our proposed Token-Level Contrastive Learning with Modality-Aware Prompting (TCL-MAP) method. As illustrated in Figure \ref{TCL_MAP}, the framework comprises two components: Prompt-Based Augmentation and Token-Level Contrastive Learning \& Intent Recognition. The former is presented following the order of Feature Extraction, Modality-Aware Prompting, and Augmented Sample Construction while the latter elucidated from the dual perspectives of Token-Level Contrastive Learning and Intent Recognition.

\subsection{Prompt-Based Augmentation}
\begin{figure*}[t!]
	\centering  
	\includegraphics[scale=.21]{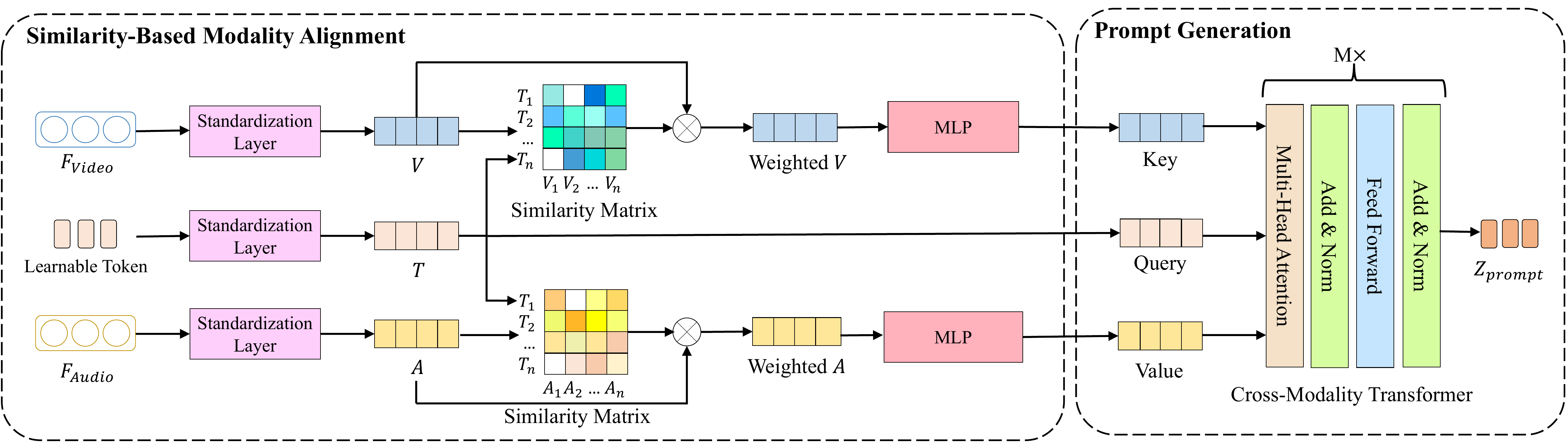}
	\caption{\protect \label{MAP} The details of Modality-Aware Prompting (MAP) module. We align multimodal features based on the content by computing the similarity matrix as weights and enhance correlations between modalities through a cross-modality transformer to create the modality-aware prompt.
}
\end{figure*}
\subsubsection{Feature Extraction}
For the text modality, we align with established practices in text intent recognition methods \cite{10349963, 10097558}, using BERT \cite{devlin2018bert}, a powerful pre-trained language model, to extract the features. To capitalize on the fine-tuning process for mining semantic information from augmented samples in later stages, we exclusively rely on the embedding layer to extract features from the text modality at this step. Specifically, Given an input utterance $t$, we get all the token representations $\textbf{Z}_{text} = [\text{CLS}, \textbf{z}_{1}, \dots, \textbf{z}_{l_{t}}] \in \mathbb{R}^{(l_{t}+1) \times d_{t}}$ from the embedding layer of BERT:
\begin{equation}
    \textbf{Z}_{text} = \text{BERTEmbedding}(t),
\end{equation}
where $\textbf{Z}$ denotes the token list, $\textbf{z}_{i}$ is the $i^{\text{th}}$ token, $CLS$ is the vector for text classification, $l_{t}$ is the sequence length of text and $d_{t}$ is the embedding size.

To extract video features, we employ a representative image classification model, Swin-Transformer \cite{Liu_2021_ICCV}, which is pre-trained on ImageNet \cite{5206848}. Concretly, we begin by dividing the raw video frames and get rid of   $[f_{1}, f_{2}, \dots, f_{l_{v}}]$. Subsequently, we obtain video feature $\textbf{F}_{video} \in \mathbb{R}^{l_{v} \times d_{v}}$ by processing each frame $f_{i}$ through Swin-Transformer to extract the feature from the last hidden layer:
\begin{equation}
    \textbf{F}_{video} = \text{Swin-Transformer}([f_{1}, f_{2}, \dots, f_{l_{v}}]),
\end{equation}
where $f_{i}$ denotes the $i^{\text{th}}$ frame, $l_{v}$ is the number of frames and $d_{v}$ is the video feature dimension. $\textbf{F}_{video}$ is composed by concatenating all individual frame features $f_{i}$.

Ultimately, we utilize a well-established pre-trained speech recognition model, Wav2Vec 2.0 \cite{NEURIPS2020_92d1e1eb}, to take the output of the last hidden layer as the audio feature $\textbf{F}_{audio} \in \mathbb{R}^{l_{a} \times d_{a}}$ for each audio segment $a$:
\begin{equation}
    \textbf{F}_{audio} = \text{Wav2Vec 2.0}(a),
\end{equation}
where $l_{a}$ is the sequence length and $d_{a}$ is the audio feature dimension.

\subsubsection{Modality-Aware Prompting}

Prior researches \cite{zhou2022learning,zhou2022conditional,rao2022denseclip} have empirically demonstrated the effectiveness of prompt learning with multimodal data, yet they tend to ignore the inherent correlations among different modalities. Inspired by this, we propose the Modality-Aware Prompting (MAP) module which integrates text, video and audio features comprehensively to attain a premium prompt representation. As shown in Figure \ref{MAP}, MAP consists of two steps: Similarity-Based Modality Alignment and Prompt Generation.

 In the Similarity-Based Modality Alignment step, following \cite{zhou2022conditional,rao2022denseclip,10.1145/1143844.1143891},we initially employ $D$ learnable tokens to approximate the actual text prompts, aiming at avoiding the substantial burden of prompt engineering and integrating multimodal information into the prompt. Then we employ individual standardization layers for each modality's feature, which incorporates a CTC \cite{10.1145/1143844.1143891} module for length normalization and an MLP to achieve consistent feature dimensions. Denoting the learnable tokens, video feature and audio feature as $\textbf{F}_{token}$, $\textbf{F}_{video}$ and $\textbf{F}_{audio}$, the process can be formulated as:
\begin{equation}
    \{\textbf{T}, \textbf{V}, \textbf{A}\} = \text{MLP}(\text{CTC}(\{\textbf{F}_{token}, \textbf{F}_{video}, \textbf{F}_{audio}\})),
\end{equation}
where $\textbf{T}, \textbf{V}, \textbf{A} \in \mathbb{R}^{L \times H}$ denotes the standardized features with length $L$ and dimension $H$. Utilizing features within the same space, we compute two similarity matrixs $\textbf{M}_{TV}, \textbf{M}_{TA} \in \mathbb{R}^{L \times L}$ based on the dot product results of normalized vectors:
\begin{equation}
    \textbf{M}_{TV} = \alpha_{TV} \cdot (\frac{\textbf{T}}{\max \limits_{i} ||T_{i}||_{2}}) (\frac{\textbf{V}^{\text{T}}}{\max \limits_{i} ||V_{i}||_{2}}),
\end{equation}
\begin{equation}
    \textbf{M}_{TA} = \alpha_{TA} \cdot (\frac{\textbf{T}}{\max \limits_{i} ||T_{i}||_{2}}) (\frac{\textbf{A}^{\text{T}}}{\max \limits_{i} ||A_{i}||_{2}}),
\end{equation}
where $\alpha_{TV}$, $\alpha_{TA}$ are threshold hyper-parameters and matrix elements $\textbf{M}_{TV}^{(ij)}$ denotes the similarity between $\textbf{T}_{i}$ and $\textbf{T}_{j}$. The matrix is subsequently subjected to a softmax activation to identify crucial features, serving as weights to reduce the discrepancy between nonverbal modalities and the learnable tokens:
\begin{equation}
    \hat{\textbf{V}} = \text{MLP}(\text{SoftMax}(\textbf{M}_{TV})\textbf{V}), 
\end{equation}
\begin{equation}
    \hat{\textbf{A}} = \text{MLP}(\text{SoftMax}(\textbf{M}_{TA})\textbf{A}),
\end{equation}
where $\hat{\textbf{V}}, \hat{\textbf{A}} \in \mathbb{R}^{L \times H}$ denotes the aligned vectors and we designate $\hat{\textbf{T}} = \textbf{T}$.

During the Prompt Generation step, we leverage cross-modality attention mechanism to effectively fuse features from three modalities into a modality-aware prompt $\textbf{Z}_{prompt}$. Following \cite{tsai2019multimodal}, We take $\hat{\textbf{T}}$ as the query, $\hat{\textbf{V}}$ as the key, and $\hat{\textbf{A}}$ as the value for the input of the multi-head attention. The attention of the $i^{\text{th}}$ head is calculated as follows:

\begin{equation}
    \text{Attention}_{i}(\textbf{Q}_{i}, \textbf{K}_{i}, \textbf{V}_{i}) = \text{SoftMax}(\frac{\textbf{Q}_{i}\textbf{K}_{i}^{\text{T}}}{\sqrt{d_{k}}})\textbf{V}_{i},
\end{equation}
where $\textbf{Q}_{i}, \textbf{K}_{i}, \textbf{V}_{i}$ are projected inputs and $d_{k}$ denotes the feature dimension $H$. The attentions of all the heads are concatenated and sequentially processed through an Add\&Norm layer, a Feed Forward layer, and another Add\&Norm layer to produce the ultimate outputs, which constitute the modality-aware prompts $\textbf{Z}_{prompt}$.

\begin{table*}[t!]
	\centering
	\begin{tabular}{@{\extracolsep{11pt}}l|cccc|cccc}
		\toprule
		\multirow{2}{*}{Methods}
            & \multicolumn{4}{c|}{MIntRec}& \multicolumn{4}{c}{MELD-DA}\\
            
		& ACC (\%) & WF1 (\%) & WP (\%) & R (\%) & ACC (\%) & WF1 (\%) & WP (\%) & R (\%) \\ 		
		\midrule
		MAG-BERT & 72.65 & 72.16 & 72.53 & 69.28 & 60.63 & 59.36 & 59.80  & 50.01  \\

            MISA & 72.29 & 72.38 & 73.48 & 69.24 & 59.98 & 58.52 & 59.28  & 48.75  \\
                 
		MulT & 72.52 & 72.31 & 72.85 & 69.24 & 60.36 & 59.01 & 59.44  & 49.93  \\
		
		\midrule
  
		TCL-MAP &
		\textbf{73.62} & 
		\textbf{73.31} & 
		\textbf{73.72} & 
		\textbf{70.50} & 
		\textbf{61.75} & 
		\textbf{59.77} & 
		\textbf{60.33}  & 
		\textbf{50.14}  \\
		$\Delta$ & 0.97$\uparrow$ & 0.93$\uparrow$ & 0.24$\uparrow$ &
		1.22$\uparrow$ & 1.12$\uparrow$ & 0.41$\uparrow$ & 0.53$\uparrow$  & 0.13$\uparrow$ \\
		\bottomrule 
	\end{tabular}
 \caption{\protect \label{results}  
	Multimodal intent recognition results on the MIntRec dataset and the MELD-DA dataset. $\Delta$ represents the maximum enhancement attained by our method compared to the baseline across the evaluation metrics.}

\end{table*} 

\begin{table*}[t!]
\renewcommand\tabcolsep{3pt} 
\centering

 \begin{tabular*}{\textwidth}{c|c@{\extracolsep{\fill}}cccccccccccc} 
 \toprule
\multicolumn{1}{l|}{Methods} &Complain& Praise & Apologise & Thank & Criticize  & Care & Agree & Taunt & Flaunt & Oppose & Joke   \\ 
\midrule
 \multicolumn{1}{l|}{MAG-BERT}  & 67.65 & 86.03 & 97.76 & 96.52 & 49.02 & 85.59 & 91.60 & 15.78 & 47.09 & 33.97 & 37.54 \\
  \multicolumn{1}{l|}{MISA}  & 63.91 & 86.63 & 97.78 & 98.03 & 53.44 & 87.14 & 92.05 & 22.15 & 46.44 & 36.15 & 38.74 \\
 \multicolumn{1}{l|}{MulT} & 65.48 & 84.72 & 97.93 & 96.83 & 49.72& 88.12 & 92.23 & 26.12 & 48.91 & 34.68 & 33.95 \\
  \multicolumn{1}{l|}{TCL-MAP}  & \textbf{68.70} & \textbf{87.20} & 97.70 & \underline{97.00} & \underline{51.30} & 86.80 & \textbf{93.10} & 17.20 & \textbf{50.80} & \underline{35.90} & 29.00 \\
  \midrule
 \multicolumn{1}{l|}{Human}  & 80.08 & 93.44 & 96.15 & 96.90 & 72.21 & 96.09 & 87.21 & 65.55 & 78.10 & 69.04 & 72.22 \\
 \midrule
\end{tabular*}

\begin{tabular*}{\textwidth}{c|c@{\extracolsep{\fill}}cccccccccc} 
 \midrule
\multicolumn{1}{l|}{Methods} & Inform  &Advise & Arrange & Introduce & Comfort & Leave & Prevent & Greet & Ask for help\\
\midrule
 \multicolumn{1}{l|}{MAG-BERT}  & 71.00 & 69.30 & 63.82 & 67.42 & 76.43 & 75.77 & 85.07 & 91.06 & 64.44 \\
  \multicolumn{1}{l|}{MISA}  & 70.18 & 69.56 & 67.32 & 67.22 & 78.78 & 77.23 & 83.30 & 82.71 & 67.57 \\
 \multicolumn{1}{l|}{MulT} & 70.85  & 69.43 & 65.44 & 71.19 & 76.44 & 75.58 & 81.68 & 86.65 & 69.12  \\
\multicolumn{1}{l|}{TCL-MAP}  & \textbf{72.80} & 68.90 & 65.40 & \underline{68.40} & \textbf{79.80} & \textbf{83.40} & \underline{83.60} & \underline{90.10} & 66.40 \\
\midrule
 \multicolumn{1}{l|}{Human}  & 79.69 & 87.14 & 81.40 & 84.09 & 95.95 & 97.06 & 86.43 & 94.15 & 88.54 \\
\bottomrule
 \end{tabular*}

\caption{\protect \label{results_each_classes} F1-Score (\%) comparison between baselines and our method for each class of MIntRec. For the results of our method, bold indicates the best performance while underlining indicates the second best performance within each class.
}
\end{table*}

\subsubsection{Augmented Sample Pair Construction}
It has been shown the superiority of text modality in achieving expressive representations for intent recognition \cite{zhang2021deep}. Inspired by this, we propose a new data augmentation method, which obtains representations of the ground truth intent labels within the semantic space provided by the text modality. Furthermore, to impose additional constraints by establishing an optimal multimodal semantic environment, we employ the modality-aware prompt to influence the label token $\textbf{z}_{label}$. To sum up, the augmented sample $\widetilde{\textbf{Z}}$ is formed by concatenating the original text tokens $\textbf{Z}_{text}$, the modality-aware prompt tokens $\textbf{Z}_{prompt}$, and the ground truth label token $\textbf{z}_{label}$, while the normal sample $\textbf{Z}$ replaces $\textbf{z}_{label}$ with the [MASK] token $\textbf{z}_{mask}$:
\begin{equation}
    \widetilde{\textbf{Z}} = \textbf{Z}_{text} \oplus \textbf{Z}_{prompt} \oplus [\textbf{z}_{label}],
\end{equation}
\begin{equation}
    \textbf{Z} = \textbf{Z}_{text} \oplus \textbf{Z}_{prompt} \oplus [\textbf{z}_{mask}],
\end{equation}
where $\oplus$ denotes the concatenation operation.

\subsection{Representation Learning}
\subsubsection{Token-Level Contrastive Learning}
To refine the normal sample $\textbf{Z}$, we utilize a powerful multimodal fusion layer MAG-BERT \cite{rahman2020integrating} to incorporate information from nonverbal modalities. For the augmented sample $\widetilde{\textbf{Z}}$, we sorely leverage the encoder layer of BERT \cite{devlin2018bert} to ensure the stability of textual semantics. After the refinement, we extract $\textbf{z}_{label}$ and $\textbf{z}_{mask}$ from their respective positions to construct the pair $(\textbf{z}_{i}, \textbf{z}_{j})$ for each sample and employ the NT-Xent \cite{sohn2016improved} loss to enhance the similarity estimation within the semantic space. This involves bringing tokens from the same pair closer while pushing apart tokens that do not belong to the same pair. Assuming $N$ represents the batch size, we can obtain a total of 2N tokens using the aforementioned approach. The contrastive loss is computed by:
\begin{equation}
    l_{ij}=-\log \frac{\exp \left(\operatorname{sim}\left(\textbf{z}_i, \textbf{z}_j\right) / \tau\right)}{\sum_{k=1}^{2 N} 1_{[k \neq i]} \exp \left(\operatorname{sim}\left(\textbf{z}_i, \textbf{z}_k\right) / \tau\right)},
\end{equation}
\begin{equation}
    \mathcal{L}_{con} = -\frac{1}{2N}\sum_{i,j}(l_{ij}+l_{ji}),
\end{equation}
where $1_{[k \neq i]}$ is an indicator function evaluating to 1 iff $k \neq i$, $\operatorname{sim}(\cdot, \cdot)$ denotes the cosine similarity between two vectors and $\tau$ denotes the temperature hyper-parameter.
\subsubsection{Classification}
Adopting the widely-used approach, we utilize the mean-pooling feature $\bar{\textbf{z}}$ of the normal tokens $Z$ for classification and use a standard entropy loss to optimize the framework:
\begin{equation}
    \mathcal{L}_{cls}=-\frac{1}{N}\sum_{i=1}^{N} \log\frac{\exp(\phi(\bar{\textbf{z}})_{y_{i}})}{\sum_{j=1}^{I}\exp(\phi(\bar{\textbf{z}})_{j})},
\end{equation}
where $N$ denotes the batch size, $\phi(\cdot)$ is the classifier with a linear layer. $y_{i}$ is the label of the $i^{\text{th}}$ sample, and $I$ is the number of labels. Ultimately, The overall learning of TCL-MAP is accomplished by minimizing the following loss:
\begin{equation}
    \mathcal{L} = \mathcal{L}_{con} + \mathcal{L}_{cls}.
\end{equation}

\begin{table*}[t!]
	\centering
	\begin{tabular}{@{\extracolsep{4.5pt}}ccc|cccc|cccc}
		\toprule
		\multicolumn{3}{c|}{Modules}
            & \multicolumn{4}{c|}{MIntRec}& \multicolumn{4}{c}{MELD-DA}\\
            
		SBMA & MAP & TCL & ACC (\%) & WF1 (\%) & WP (\%) & R (\%) & ACC (\%) & WF1 (\%) & WP (\%) & R (\%) \\ 		
		\midrule
             $\scalebox{0.75}{\usym{2613}}$ & $\checkmark$ & $\checkmark$ & 73.15 & 72.73 & 73.02 & 69.82 & 61.24 & 59.32 & 59.58  & 49.89  \\
             $\scalebox{0.75}{\usym{2613}}$ & $\scalebox{0.75}{\usym{2613}}$ & $\checkmark$  & 72.67 & 72.31 & 72.81 & 69.78 & 60.40 & 58.69 & 59.78  & 49.63  \\

             $\checkmark$ & $\checkmark$ & $\scalebox{0.75}{\usym{2613}}$  & 72.13 & 71.80 & 72.50 & 68.80 & 61.26 & 59.54 & 59.97  & 50.09  \\
             \midrule
             $\checkmark$ & $\checkmark$ & $\checkmark$ &
		\textbf{73.62} & 
		\textbf{73.31} & 
		\textbf{73.72} & 
		\textbf{70.50} & 
		\textbf{61.75} & 
		\textbf{59.77} & 
		\textbf{60.33}  & 
		\textbf{50.14}  \\
		\bottomrule 
	\end{tabular}
 \caption{\protect \label{ablation}  
	Ablation experiments of modules in TCL-MAP on the MIntRec dataset and the MELD-DA dataset. SBMA stands for Similarity-Based Modality Alignment, MAP stands for Modality-Aware Prompt and TCL stands for Token-Level Contrastive Learning. With SBMA incorporated into MAP, there exist three distinct settings.}

\end{table*}

\section{Experiments}
\subsection{Datasets}
We conduct experiments on two challenging multimodal datasets to evaluate our proposed framework.
\subsubsection{MIntRec}
MIntRec \cite{10.1145/3503161.3547906} is a fine-grained dataset for multimodal intent recognition with 2,224 high-quality samples with text, video and audio modalities across 20 intent categories. We follow the dataset splits consisting of 1,334 samples for training, 445 samples for validation, and 445 samples for testing.

\subsubsection{MELD-DA}
MELD-DA \cite{saha-etal-2020-towards} is a large-scale dataset for dialogue act classification, comprising 9988 multimodal samples. The dataset is divided into a training set of 6,991 samples, a validation set of 999 samples and a test set of 1,998 samples. All data are annotated with 12 common dialogue act labels.

\subsection{Baselines}
Following \cite{10.1145/3503161.3547906}, we use the following state-of-the-art multimodal fusion methods as the baselines: (1) MAG-BERT \cite{rahman2020integrating} presents an efficient attachment for pre-trained language models to expand the capabilities to fine-tune on multimodal representations; (2) MulT \cite{tsai2019multimodal} utilizes directional pairwise cross-modality attention without explicit alignment to address interactions between multimodal sequences; (3) MISA \cite{10.1145/3394171.3413678} captures both modality-invariant and modality-specific features from each modality and subsequently fuses them with self-attention mechanism.

\subsection{Evaluation Metrics}
We adopt accuracy (ACC), weighted F1-score (WF1), weighted precision (WP) and recall (R) to evaluate the model performance, which is commonly used in classification tasks. Considering the imbalance across different categories, for the second and the third metrics, we present the scores weighted by the number of samples in each class. Higher values indicate improved performance across all metrics.

\subsection{Experimental Settings}
For the implementation of our proposed method, we utilize bert-base-uncased and wav2vec2-base-960h from Huggingface Library \cite{wolf2019huggingface} to extract text and audio features and swin\_b pre-trained on ImageNet1K \cite{5206848} from Torchvision Library \cite{torchvision2016} to extract video features. The training batch size is set to 16, while the validation and test batch sizes are both set to 8. For the total loss $\mathcal{L}$, we employ AdamW \cite{loshchilov2017decoupled} to optimize the parameters.

\subsection{Results}
We conduct experiments on both the MIntRec dataset and the MELD-DA dataset, comparing our approach with state-of-the-art baselines. The results are presented in Table \ref{results} with the optimal outcomes highlighted in bold, and the enhancements of our method over the top-performing baseline are indicated by $\Delta$.

Firstly, we observe the overall performance. As indicated by the results, our approach has consistently outperformed the current state-of-the-art methods across all four metrics on both datasets, demonstrating significant advancements. Secondly, on the MIntRec dataset, our approach demonstrates enhancements of 0.97\% on ACC, 0.93\% on WF1 and 1.22\% on R, which indicates the robust capability of our approach to effectively leverage multimodal information for the extraction and identification of intricate intents in real-world scenarios. Thirdly, on the MELD-DA dataset, our method also achieves notable improvements on both ACC and WP metrics, despite the presence of challenging ``Others'' label which is difficult to distinguish. This observation showcases the effectiveness of our method in recognizing ambiguous intents such as dialogue acts.

\section{Discussion}
\subsection{Effectiveness of Each Module}
To further analyze the individual contributions of the modules within TCL-MAP to the overall performance, we conducted the following ablation experiments and the results are illustrated in Table \ref{ablation}.

\subsubsection{Similarity-Based Modality Alignment}
To assess the effectiveness of similarity-based modality alignment, we replace the alignment method with a CTC module \cite{10.1145/1143844.1143891} which aligns multimodal features solely from a temporal perspective and disregards the correlations. As indicated by the results, the performance of TCL-MAP exhibites a reduction of more than 0.50\% across most metrics for both datasets. The most significant decrease, amounting to 0.75\%, is observed in the WP metric for the MELD-DA dataset. These observations illustrate the effectiveness of our proposed similarity-based modality alignment in aligning multimodal features and facilitating the extraction of semantic information. Moreover, even without the presence of similarity-based modality alignment, TCL-MAP continues to achieve superior results on the MIntRec dataset, underscoring the efficacy of the other modules.

\subsubsection{Modality-Aware Prompting}
In this setting, we remove modality-aware prompting module and directly concatenate the original text tokens with the [MASK]/Label token as the augmented pair. We note more substantial reductions on MIntRec, such as a 0.95\% decrease on ACC and a 1.00\% decrease on WF1. Meanwhile, the performance on MELD-DA experiences notable declines on ACC, WF1 and WP metrics. We attribute this to the fact that the optimal semantic environment created by our modality-aware prompting module aids in filtering out irrelevant semantics within the [MASK]/Label token, which makes the token-level contrastive learning more precise.

\subsubsection{Token-Level Contrastive Learning}
In the absence of token-level contrastive learning, we exclude the contrastive learning loss $\mathcal{L}_{con}$ from the total loss $\mathcal{L}$ and proceed with the learning process guided by classification. In this experimental setup, all of the four metrics decrease by 1.49\%, 1.51\%, 1.22\% and 1.70\% on MIntRec and the ACC metric and the WP metric decrease by 0.49\% and 0.36\% on MELD-DA, indicating a significant decline on performance. The experimental results suggest that our introduced token-level contrastive learning effectively leverages the rich semantic information within the ground truth labels to guide the learning process of nonverbal modalities and simultaneously optimizes feature representations together with the classification guidance, leading to improved performance.

\subsection{Performance of Fine-grained Intent Classes}
To examine the performance of our method in each fine-grained intent classes, we compare the F1 scores of TCL-MAP and baseline methods for each intent class in MIntRec. As shown in Table \ref{results_each_classes}, the results are obtained by averaging the scores over ten runs of experiments with different random seeds and for the scores of TCL-MAP we mark the best results in bold and the second best results with underlines within each class.

To begin with, we analyze the comprehensive results of our proposed TCL-MAP in comparison to the baseline methods. Remarkably, across all 20 intent categories, our approach attains top-2 scores in 13 categories, comprising 7 highest scores and 6 second highest scores, which indicates that  TCL-MAP achieves better performance than the majority of baselines across various classes. Specifically, in categories like``Complain'', ``Agree'' and ``Leave'', TCL-MAP consistently outperforms the best baseline by over 1\%. Significantly, the ``Leave'' category exhibits the most substantial improvement of 7.63\%. The significant gains can be attributed to TCL-MAP's utilization of modality-aware prompts for better text representation, which in turn enhances video and audio learning through token-level contrastive learning. Nevertheless, in the ``Taunt'' and ``Joke'' categories, TCL-MAP seems to provide less assistance in recognizing the intent, which could be caused by a combination of factors, including the limited availability of data within these categories and the intricate nature of the intents themselves.

On the other hand, we evaluate the efficacy of TCL-MAP in comparison to the human performance. From the results, we can observe that humans achieve the best performance in the majority of intent categories, which confirms the strong ability of humans to process multimodal information and infer intents through them. However, TCL-MAP surpasses human performance in the "Apologize," "Thank," and "Agree" classes, showcasing the stability of our method when handling challenging samples where humans may make mistakes. In addition, TCL-MAP has approached human performance in intent categories (e.g. ``complain'', ``praise'' and ``care'') which involve distinct emotional aspects and also achieved comparable performance to humans in intent categories (e.g. ``Inform'', ``leave'' and ``prevent'' ) which require an understanding of actions. These findings further validate the capability of TCL-MAP to effectively extract fearures related to human intents from raw multimodal data, such as expressions, tone of speech and movements.

\subsection{Comparison between Handcraft Prompt and Modality-Aware Prompt}
To further analyze the superiority of our modality-aware prompt, we conduct experiments with handcrafted prompt and modality-aware prompt respectively. Concretly, we select the MIntRec dataset for our experiments, driven by the fact that certain labels (e.g. ``Others'') in the MELD-DA dataset do not strictly represent intent categories. To make comparison, we design two handcraft prompts aimed at expressing ideas or intents, ``I want to'' and ``I intend to'', which maintain the same positions and lengths with the modality-aware prompt. Besides, we conduct an additional set of experiments using [MASK] as the prompt to demonstrate the effectiveness.

\begin{figure}[t!]
	\centering  
	\includegraphics[scale=.27]{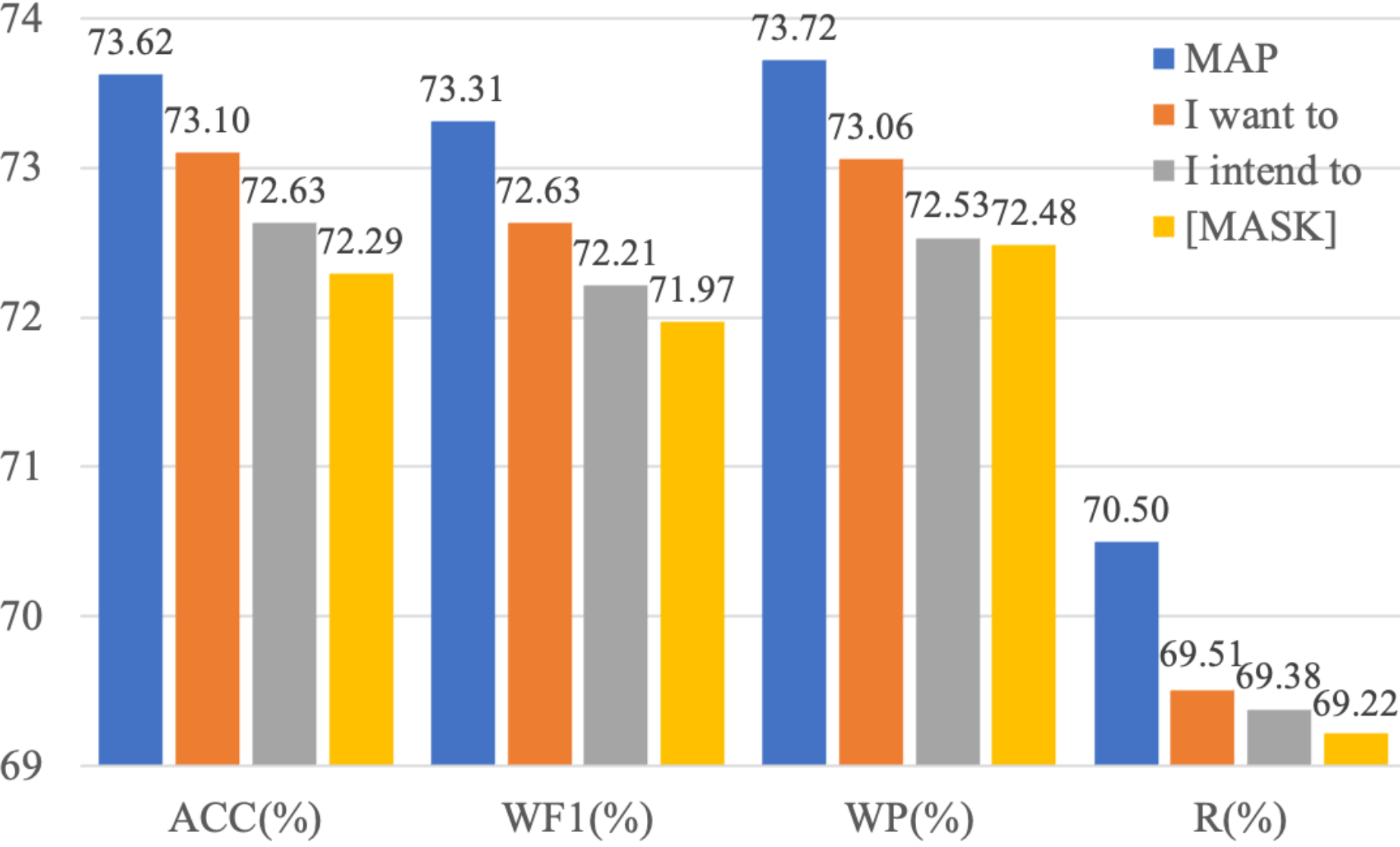}
	\caption{\protect \label{analyze} The comparison between Handcraft Prompt and Modality-Aware Prompt
 }
\end{figure}

As shown in Figure \ref{analyze}, we observe a substantial performance advantage in the model that employs the modality-aware prompt in comparison to models using handcrafted prompts, thanks to better integration of non-textual modalities enhancing textual intent semantics extraction. Conversely, the [MASK] prompt shows a notable performance decline compared to handcrafted prompts, highlighting the risk of inappropriate prompts misleading intent understanding. Our modality-aware prompt incorporates the instance-conditional prompt concept of CoCoOp \cite{zhou2022conditional}, thereby mitigating this drawback.

\section{Conclusion}
In this paper, we propose a novel Token-Level Contrastive Learning with Modality-Aware Prompting (TCL-MAP) method for multimodal intent recognition. By strengthening the correlations among modalities, our method generate the modality-aware prompt to construct an optimal multimodal semantic space for enhancing the refinement of the text modality. In return, the attained textual representation, enriched with semantics from the ground truth label token, guides the learning process of nonverbal modalities through the token-level contrastive learning. Extensive experiments on two benchmark datasets demonstrate that our approach outperforms state-of-the-art methods and carries significant implications for multimodal prompt learning.

\section*{Acknowledgements}
This work is founded by the National Natural Science Foundation of China (Grant No. 62173195), National Science and Technology Major Project towards the new generation of broadband wireless mobile communication networks of Jiangxi Province (Grant No.20232ABC03402), High-level Scientific and Technological Innovation Talents ``Double Hundred Plan'' of Nanchang City (Grant No. Hongke Zi (2022) 321-16), and Natural
Science Foundation of Hebei Province, China (Grant No. F2022208006).

\bibliography{aaai24}

\begin{thebibliography}{42}
\providecommand{\natexlab}[1]{#1}

\bibitem[{Baevski et~al.(2020)Baevski, Zhou, Mohamed, and Auli}]{NEURIPS2020_92d1e1eb}
Baevski, A.; Zhou, Y.; Mohamed, A.; and Auli, M. 2020.
\newblock wav2vec 2.0: A Framework for Self-Supervised Learning of Speech Representations.
\newblock In Larochelle, H.; Ranzato, M.; Hadsell, R.; Balcan, M.; and Lin, H., eds., \emph{Advances in Neural Information Processing Systems}, volume~33, 12449--12460. Curran Associates, Inc.

\bibitem[{Caron et~al.(2021)Caron, Touvron, Misra, J{\'e}gou, Mairal, Bojanowski, and Joulin}]{caron2021emerging}
Caron, M.; Touvron, H.; Misra, I.; J{\'e}gou, H.; Mairal, J.; Bojanowski, P.; and Joulin, A. 2021.
\newblock Emerging properties in self-supervised vision transformers.
\newblock In \emph{Proceedings of the IEEE/CVF international conference on computer vision}, 9650--9660.

\bibitem[{Chen et~al.(2020)Chen, Kornblith, Norouzi, and Hinton}]{chen2020simple}
Chen, T.; Kornblith, S.; Norouzi, M.; and Hinton, G. 2020.
\newblock A simple framework for contrastive learning of visual representations.
\newblock In \emph{International conference on machine learning}, 1597--1607. PMLR.

\bibitem[{Chen and He(2021)}]{chen2021exploring}
Chen, X.; and He, K. 2021.
\newblock Exploring simple siamese representation learning.
\newblock In \emph{Proceedings of the IEEE/CVF conference on computer vision and pattern recognition}, 15750--15758.

\bibitem[{Deng et~al.(2009)Deng, Dong, Socher, Li, Li, and Fei-Fei}]{5206848}
Deng, J.; Dong, W.; Socher, R.; Li, L.-J.; Li, K.; and Fei-Fei, L. 2009.
\newblock ImageNet: A large-scale hierarchical image database.
\newblock In \emph{2009 IEEE Conference on Computer Vision and Pattern Recognition}, 248--255.

\bibitem[{Devlin et~al.(2018)Devlin, Chang, Lee, and Toutanova}]{devlin2018bert}
Devlin, J.; Chang, M.-W.; Lee, K.; and Toutanova, K. 2018.
\newblock Bert: Pre-training of deep bidirectional transformers for language understanding.
\newblock \emph{arXiv preprint arXiv:1810.04805}.

\bibitem[{Dong et~al.(2022)Dong, Fu, Zhou, Li, and Wang}]{dong2022improving}
Dong, J.; Fu, J.; Zhou, P.; Li, H.; and Wang, X. 2022.
\newblock Improving spoken language understanding with cross-modal contrastive learning.
\newblock \emph{Interspeech. ISCA}.

\bibitem[{Gan et~al.(2023)Gan, Bai, Lou, Ma, Zhang, Shi, and Luo}]{gan2023decorate}
Gan, Y.; Bai, Y.; Lou, Y.; Ma, X.; Zhang, R.; Shi, N.; and Luo, L. 2023.
\newblock Decorate the newcomers: Visual domain prompt for continual test time adaptation.
\newblock In \emph{Proceedings of the AAAI Conference on Artificial Intelligence}, volume~37, 7595--7603.

\bibitem[{Graves et~al.(2006)Graves, Fern\'{a}ndez, Gomez, and Schmidhuber}]{10.1145/1143844.1143891}
Graves, A.; Fern\'{a}ndez, S.; Gomez, F.; and Schmidhuber, J. 2006.
\newblock Connectionist Temporal Classification: Labelling Unsegmented Sequence Data with Recurrent Neural Networks.
\newblock In \emph{Proceedings of the 23rd International Conference on Machine Learning}, ICML '06, 369–376. New York, NY, USA: Association for Computing Machinery.
\newblock ISBN 1595933832.

\bibitem[{Grill et~al.(2020)Grill, Strub, Altch{\'e}, Tallec, Richemond, Buchatskaya, Doersch, Avila~Pires, Guo, Gheshlaghi~Azar et~al.}]{grill2020bootstrap}
Grill, J.-B.; Strub, F.; Altch{\'e}, F.; Tallec, C.; Richemond, P.; Buchatskaya, E.; Doersch, C.; Avila~Pires, B.; Guo, Z.; Gheshlaghi~Azar, M.; et~al. 2020.
\newblock Bootstrap your own latent-a new approach to self-supervised learning.
\newblock \emph{Advances in neural information processing systems}, 33: 21271--21284.

\bibitem[{Han et~al.(2021)Han, Chen, Gelbukh, Zadeh, Morency, and Poria}]{han2021bi}
Han, W.; Chen, H.; Gelbukh, A.; Zadeh, A.; Morency, L.-p.; and Poria, S. 2021.
\newblock Bi-bimodal modality fusion for correlation-controlled multimodal sentiment analysis.
\newblock In \emph{Proceedings of the 2021 International Conference on Multimodal Interaction}, 6--15.

\bibitem[{Han, Chen, and Poria(2021)}]{han2021improving}
Han, W.; Chen, H.; and Poria, S. 2021.
\newblock Improving multimodal fusion with hierarchical mutual information maximization for multimodal sentiment analysis.
\newblock \emph{arXiv preprint arXiv:2109.00412}.

\bibitem[{Hazarika, Zimmermann, and Poria(2020)}]{10.1145/3394171.3413678}
Hazarika, D.; Zimmermann, R.; and Poria, S. 2020.
\newblock MISA: Modality-Invariant and -Specific Representations for Multimodal Sentiment Analysis.
\newblock In \emph{Proceedings of the 28th ACM International Conference on Multimedia}, MM '20, 1122–1131. New York, NY, USA: Association for Computing Machinery.
\newblock ISBN 9781450379885.

\bibitem[{He et~al.(2020)He, Fan, Wu, Xie, and Girshick}]{he2020momentum}
He, K.; Fan, H.; Wu, Y.; Xie, S.; and Girshick, R. 2020.
\newblock Momentum contrast for unsupervised visual representation learning.
\newblock In \emph{Proceedings of the IEEE/CVF conference on computer vision and pattern recognition}, 9729--9738.

\bibitem[{Hou et~al.(2019)Hou, Tang, Zhang, Kong, and Zhao}]{hou2019deep}
Hou, M.; Tang, J.; Zhang, J.; Kong, W.; and Zhao, Q. 2019.
\newblock Deep multimodal multilinear fusion with high-order polynomial pooling.
\newblock \emph{Advances in Neural Information Processing Systems}, 32.

\bibitem[{Li et~al.(2022)Li, Li, Li, Niebles, and Hoi}]{Li_2022_CVPR}
Li, D.; Li, J.; Li, H.; Niebles, J.~C.; and Hoi, S.~C. 2022.
\newblock Align and Prompt: Video-and-Language Pre-Training With Entity Prompts.
\newblock In \emph{Proceedings of the IEEE/CVF Conference on Computer Vision and Pattern Recognition (CVPR)}, 4953--4963.

\bibitem[{Liu et~al.(2021)Liu, Lin, Cao, Hu, Wei, Zhang, Lin, and Guo}]{Liu_2021_ICCV}
Liu, Z.; Lin, Y.; Cao, Y.; Hu, H.; Wei, Y.; Zhang, Z.; Lin, S.; and Guo, B. 2021.
\newblock Swin Transformer: Hierarchical Vision Transformer Using Shifted Windows.
\newblock In \emph{Proceedings of the IEEE/CVF International Conference on Computer Vision (ICCV)}, 10012--10022.

\bibitem[{Liu et~al.(2018)Liu, Shen, Lakshminarasimhan, Liang, Zadeh, and Morency}]{liu2018efficient}
Liu, Z.; Shen, Y.; Lakshminarasimhan, V.~B.; Liang, P.~P.; Zadeh, A.; and Morency, L.-P. 2018.
\newblock Efficient low-rank multimodal fusion with modality-specific factors.
\newblock \emph{arXiv preprint arXiv:1806.00064}.

\bibitem[{Loshchilov and Hutter(2017)}]{loshchilov2017decoupled}
Loshchilov, I.; and Hutter, F. 2017.
\newblock Decoupled weight decay regularization.
\newblock \emph{arXiv preprint arXiv:1711.05101}.

\bibitem[{maintainers and contributors(2016)}]{torchvision2016}
maintainers, T.; and contributors. 2016.
\newblock TorchVision: PyTorch's Computer Vision library.
\newblock \url{https://github.com/pytorch/vision}.
\newblock Accessed: 2023-08-15.

\bibitem[{Paraskevopoulos, Georgiou, and Potamianos(2022)}]{paraskevopoulos2022mmlatch}
Paraskevopoulos, G.; Georgiou, E.; and Potamianos, A. 2022.
\newblock Mmlatch: Bottom-up top-down fusion for multimodal sentiment analysis.
\newblock In \emph{ICASSP 2022-2022 IEEE International Conference on Acoustics, Speech and Signal Processing (ICASSP)}, 4573--4577. IEEE.

\bibitem[{Rahman et~al.(2020)Rahman, Hasan, Lee, Zadeh, Mao, Morency, and Hoque}]{rahman2020integrating}
Rahman, W.; Hasan, M.~K.; Lee, S.; Zadeh, A.; Mao, C.; Morency, L.-P.; and Hoque, E. 2020.
\newblock Integrating multimodal information in large pretrained transformers.
\newblock In \emph{Proceedings of the conference. Association for Computational Linguistics. Meeting}, volume 2020, 2359. NIH Public Access.

\bibitem[{Rao et~al.(2022)Rao, Zhao, Chen, Tang, Zhu, Huang, Zhou, and Lu}]{rao2022denseclip}
Rao, Y.; Zhao, W.; Chen, G.; Tang, Y.; Zhu, Z.; Huang, G.; Zhou, J.; and Lu, J. 2022.
\newblock Denseclip: Language-guided dense prediction with context-aware prompting.
\newblock In \emph{Proceedings of the IEEE/CVF Conference on Computer Vision and Pattern Recognition}, 18082--18091.

\bibitem[{Saha et~al.(2020)Saha, Patra, Saha, and Bhattacharyya}]{saha-etal-2020-towards}
Saha, T.; Patra, A.; Saha, S.; and Bhattacharyya, P. 2020.
\newblock Towards Emotion-aided Multi-modal Dialogue Act Classification.
\newblock In \emph{Proceedings of the 58th Annual Meeting of the Association for Computational Linguistics}, 4361--4372. Online: Association for Computational Linguistics.

\bibitem[{Sohn(2016)}]{sohn2016improved}
Sohn, K. 2016.
\newblock Improved deep metric learning with multi-class n-pair loss objective.
\newblock \emph{Advances in neural information processing systems}, 29.

\bibitem[{Tian, Krishnan, and Isola(2020)}]{tian2020contrastive}
Tian, Y.; Krishnan, D.; and Isola, P. 2020.
\newblock Contrastive multiview coding.
\newblock In \emph{Computer Vision--ECCV 2020: 16th European Conference, Glasgow, UK, August 23--28, 2020, Proceedings, Part XI 16}, 776--794. Springer.

\bibitem[{Tsai et~al.(2019)Tsai, Bai, Liang, Kolter, Morency, and Salakhutdinov}]{tsai2019multimodal}
Tsai, Y.-H.~H.; Bai, S.; Liang, P.~P.; Kolter, J.~Z.; Morency, L.-P.; and Salakhutdinov, R. 2019.
\newblock Multimodal transformer for unaligned multimodal language sequences.
\newblock In \emph{Proceedings of the conference. Association for Computational Linguistics. Meeting}, volume 2019, 6558. NIH Public Access.

\bibitem[{Wang et~al.(2022)Wang, Zhang, Lee, Zhang, Sun, Ren, Su, Perot, Dy, and Pfister}]{Wang_2022_CVPR}
Wang, Z.; Zhang, Z.; Lee, C.-Y.; Zhang, H.; Sun, R.; Ren, X.; Su, G.; Perot, V.; Dy, J.; and Pfister, T. 2022.
\newblock Learning To Prompt for Continual Learning.
\newblock In \emph{Proceedings of the IEEE/CVF Conference on Computer Vision and Pattern Recognition (CVPR)}, 139--149.

\bibitem[{Wolf et~al.(2019)Wolf, Debut, Sanh, Chaumond, Delangue, Moi, Cistac, Rault, Louf, Funtowicz et~al.}]{wolf2019huggingface}
Wolf, T.; Debut, L.; Sanh, V.; Chaumond, J.; Delangue, C.; Moi, A.; Cistac, P.; Rault, T.; Louf, R.; Funtowicz, M.; et~al. 2019.
\newblock Huggingface's transformers: State-of-the-art natural language processing.
\newblock \emph{arXiv preprint arXiv:1910.03771}.

\bibitem[{Wu et~al.(2018)Wu, Xiong, Yu, and Lin}]{wu2018unsupervised}
Wu, Z.; Xiong, Y.; Yu, S.~X.; and Lin, D. 2018.
\newblock Unsupervised feature learning via non-parametric instance discrimination.
\newblock In \emph{Proceedings of the IEEE conference on computer vision and pattern recognition}, 3733--3742.

\bibitem[{Ye et~al.(2019)Ye, Zhang, Yuen, and Chang}]{ye2019unsupervised}
Ye, M.; Zhang, X.; Yuen, P.~C.; and Chang, S.-F. 2019.
\newblock Unsupervised embedding learning via invariant and spreading instance feature.
\newblock In \emph{Proceedings of the IEEE/CVF conference on computer vision and pattern recognition}, 6210--6219.

\bibitem[{Yu et~al.(2023)Yu, Gao, Lin, Yang, Wu, Ma, Wang, Huang, and Li}]{yu-etal-2023-speech}
Yu, T.; Gao, H.; Lin, T.-E.; Yang, M.; Wu, Y.; Ma, W.; Wang, C.; Huang, F.; and Li, Y. 2023.
\newblock Speech-Text Pre-training for Spoken Dialog Understanding with Explicit Cross-Modal Alignment.
\newblock In \emph{Proceedings of the 61st Annual Meeting of the Association for Computational Linguistics (Volume 1: Long Papers)}, 7900--7913. Toronto, Canada: Association for Computational Linguistics.

\bibitem[{Zadeh et~al.(2017)Zadeh, Chen, Poria, Cambria, and Morency}]{zadeh2017tensor}
Zadeh, A.; Chen, M.; Poria, S.; Cambria, E.; and Morency, L.-P. 2017.
\newblock Tensor fusion network for multimodal sentiment analysis.
\newblock \emph{arXiv preprint arXiv:1707.07250}.

\bibitem[{Zadeh et~al.(2018)Zadeh, Liang, Mazumder, Poria, Cambria, and Morency}]{zadeh2018memory}
Zadeh, A.; Liang, P.~P.; Mazumder, N.; Poria, S.; Cambria, E.; and Morency, L.-P. 2018.
\newblock Memory fusion network for multi-view sequential learning.
\newblock In \emph{Proceedings of the AAAI conference on artificial intelligence}, volume~32.

\bibitem[{Zhang et~al.(2021{\natexlab{a}})Zhang, Li, Xu, Zhang, Zhao, and Gao}]{zhang2021textoir}
Zhang, H.; Li, X.; Xu, H.; Zhang, P.; Zhao, K.; and Gao, K. 2021{\natexlab{a}}.
\newblock TEXTOIR: An Integrated and Visualized Platform for Text Open Intent Recognition.
\newblock In \emph{Proceedings of the 59th Annual Meeting of the Association for Computational Linguistics and the 11th International Joint Conference on Natural Language Processing: System Demonstrations}, 167--174.

\bibitem[{Zhang, Xu, and Lin(2021)}]{zhang2021deep}
Zhang, H.; Xu, H.; and Lin, T.-E. 2021.
\newblock Deep open intent classification with adaptive decision boundary.
\newblock In \emph{Proceedings of the AAAI Conference on Artificial Intelligence}, volume~35, 14374--14382.

\bibitem[{Zhang et~al.(2021{\natexlab{b}})Zhang, Xu, Lin, and Lyu}]{zhang2021discovering}
Zhang, H.; Xu, H.; Lin, T.-E.; and Lyu, R. 2021{\natexlab{b}}.
\newblock Discovering new intents with deep aligned clustering.
\newblock In \emph{Proceedings of the AAAI Conference on Artificial Intelligence}, volume~35, 14365--14373.

\bibitem[{Zhang et~al.(2023{\natexlab{a}})Zhang, Xu, Wang, Long, and Gao}]{10349963}
Zhang, H.; Xu, H.; Wang, X.; Long, F.; and Gao, K. 2023{\natexlab{a}}.
\newblock A Clustering Framework for Unsupervised and Semi-supervised New Intent Discovery.
\newblock \emph{IEEE Transactions on Knowledge and Data Engineering}, 1--14.

\bibitem[{Zhang et~al.(2022)Zhang, Xu, Wang, Zhou, Zhao, and Teng}]{10.1145/3503161.3547906}
Zhang, H.; Xu, H.; Wang, X.; Zhou, Q.; Zhao, S.; and Teng, J. 2022.
\newblock MIntRec: A New Dataset for Multimodal Intent Recognition.
\newblock In \emph{Proceedings of the 30th ACM International Conference on Multimedia}, MM '22, 1688–1697. New York, NY, USA: Association for Computing Machinery.
\newblock ISBN 9781450392037.

\bibitem[{Zhang et~al.(2023{\natexlab{b}})Zhang, Xu, Zhao, and Zhou}]{10097558}
Zhang, H.; Xu, H.; Zhao, S.; and Zhou, Q. 2023{\natexlab{b}}.
\newblock Learning Discriminative Representations and Decision Boundaries for Open Intent Detection.
\newblock \emph{IEEE/ACM Transactions on Audio, Speech, and Language Processing}, 31: 1611--1623.

\bibitem[{Zhou et~al.(2022{\natexlab{a}})Zhou, Yang, Loy, and Liu}]{zhou2022conditional}
Zhou, K.; Yang, J.; Loy, C.~C.; and Liu, Z. 2022{\natexlab{a}}.
\newblock Conditional prompt learning for vision-language models.
\newblock In \emph{Proceedings of the IEEE/CVF Conference on Computer Vision and Pattern Recognition}, 16816--16825.

\bibitem[{Zhou et~al.(2022{\natexlab{b}})Zhou, Yang, Loy, and Liu}]{zhou2022learning}
Zhou, K.; Yang, J.; Loy, C.~C.; and Liu, Z. 2022{\natexlab{b}}.
\newblock Learning to prompt for vision-language models.
\newblock \emph{International Journal of Computer Vision}, 130(9): 2337--2348.

\end{thebibliography}

\end{document}